\documentclass{article}
\usepackage[utf8]{inputenc}
\usepackage{color}
\usepackage{array}
\usepackage{url}
\usepackage{amsmath}
\usepackage{amsfonts}
\usepackage{graphicx}
\usepackage{esint}
\usepackage[numbers]{natbib}
\usepackage[unicode=true,pdfusetitle,bookmarks=true,bookmarksnumbered=false,bookmarksopen=false,
breaklinks=true,pdfborder={0 0 1},backref=false,colorlinks=true]{hyperref}
\hypersetup{linkcolor=blue,citecolor=blue}

\makeatletter

\providecommand{\tabularnewline}{\\}

\usepackage{authblk}
\usepackage{calc}
\usepackage{gensymb}
\usepackage{listings}
\usepackage[numbered,framed]{mcode}
\usepackage[space]{grffile}%% hyperref manual advises loading this package last
\usepackage{lineno}
\usepackage{xcolor}
\usepackage{textcomp}

\definecolor{gray}{rgb}{0.9,0.9,0.9}
\makeatother

\begin{document}

\title{LCS Tool : A Computational Platform for Lagrangian Coherent Structures}
\author[1]{K. Onu}
\author[2]{F. Huhn}
\author[2]{G. Haller}
\affil[1]{Mechanical Engineering, McGill University, Montreal, Canada.}
\affil[2]{Institute of Mechanical Systems, ETH Zurich, Switzerland.}
\maketitle
\begin{abstract}
We give an algorithmic introduction to Lagrangian coherent structures
(LCSs) using a newly developed computational engine, LCS Tool. LCSs
are most repelling, attracting and shearing material lines that form
the centerpieces of observed tracer patterns in two-dimensional unsteady
dynamical systems. LCS Tool implements the latest geodesic theory
of LCSs for two-dimensional flows, uncovering key transport barriers
in unsteady flow velocity data as explicit solutions of differential
equations. After a review of the underlying theory, we explain the
steps and numerical methods used by LCS Tool, and illustrate its capabilities
on three unsteady fluid flow examples.
\end{abstract}

\section{Introduction}
Lagrangian Coherent Structures (LCSs) are evolving organizing centers
of trajectory patterns in non-autonomous dynamical systems \citep{Haller2000,peacock13:_lagran,Haller2014}.
Applications of LCSs include oceanic and atmospheric flows \citep{beron-vera13:_objec_agulh,Koh2002},
biological transport problems \citep{wilson09:_lagran_reynol,tallapragada11:_lagran,Huhn2012},
aeronautics \citep{tang10:_accur_lagran_hong_kong_inter_airpor},
celestial mechanics \citep{Gawlik2009}, crowd dynamics \citep{Ali2007},
and aperiodically forced mechanical oscillators \citep{hadjighasem13:_detec_kam}.

\citet{Haller2001} proposed that ridges of the Finite-Time Lyapunov
exponent (FTLE) are heuristic indicators of hyperbolic (i.e., repelling-
and attracting-type) LCSs. A number of examples indeed support this
principle \citep{peacock13:_lagran}. Equating
FTLE ridges with LCSs, however, would create theoretical inconsistencies,
as well as false positives and negatives in hyperbolic LCS detection
\citep{haller11:_lagran_coher_struc,norgard12:_secon_lagran_coher_struc}.
In addition, the role of the FTLE field in the accurate detection
of elliptic (vortex-type) and parabolic (jet-core type) LCSs has remained
an open question (but see \citep{beron-vera10:_invar_lagran}).

More recent work has focused on an exact mathematical formulation
of the properties defining LCSs \citep{haller11:_lagran_coher_struc,farazmand12:_comput_lagran,haller12:_geodes_theor_trans_barrier_two_dimen_flows,haller13:_coher_lagran,Blazevski2014}.
In two-dimensional flows, hyperbolic and parabolic LCSs turn out to
be stationary curves of the averaged material shear \citep{Farazmand2014},
whereas elliptic LCSs are stationary curves of the averaged strain
\citep{haller13:_coher_lagran}. These variational formulations lead
to explicit solutions for LCSs as null-geodesics of appropriate Lorentzian
metrics. 

Here we present a simple algorithmic introduction to geodesic LCS
detection without details on the the underlying mathematics. We then
review the implementation of this approach in a computational engine
called LCS Tool%
\footnote{LCS Tool is publicly available at: \emph{\url{https://github.com/jeixav/LCS-Tool}}.%
}. This engine is a library of MATLAB functions that extract LCSs from
two-dimensional unsteady flows. The examples we present here have
been programmed into demonstration scripts.

The list of publicly available software for LCS detection further
includes the following packages:
\begin{itemize}
\item \emph{MANGEN} \citep{lekien03:_time} calculates FTLE and advects
material curves in two-dimensional velocity fields. It also includes
a graphical user interface and uses MPI for parallel calculations. 
\item \emph{LCS MATLAB Kit} \citep{dabiri09:_lmk} calculates the FTLE from
velocity datasets. Includes a graphical user interface. 
\item \emph{Newman} \citep{toit10:_trans} calculates the FTLE in $N$ dimensions.
Assists ridge extraction of FTLE fields. Supports analytic- and dataset-based
velocity definitions. 
\item \emph{FlowVC} \citep{shadden10:_flowvc} is a general-purpose LCS
platform for two- and three- dimensional datasets. Parallel calculations
are supported with Open\-MP, CUDA and OpenCL. 
\end{itemize}
These packages primarily target the automated generation of FTLE plots
to aid the visual assessment of hyperbolic LCSs in the flow. The purpose
of LCS Tool and this article is to provide an all-purpose numerical
engine for the geodesic extraction of LCSs as parametrized material
curves in unsteady two-dimensional flows, and extend the scope of
such extraction to elliptic LCSs.

\section{Theory}
We consider two-dimensional, finite-time, unsteady velocity fields
of the form 
\begin{align}
\dot{x}=v(x,t), &  & x\in U\subset\mathbb{R}^{2}, &  & t\in[t_{-},t_{+}].\label{eq:v(x,t)}
\end{align}
Trajectories of \eqref{eq:v(x,t)} are denoted $x(t;t_{0},x_{0})$,
with $x_{0}\in U$ denoting their initial position in the open set
$U$ at an initial time $t_{0}\in[t_{-},t_{+}]$. The flow map is
then defined as 
\[
F_{t_{0}}^{t}(x_{0}):=x(t;t,x_{0}),
\]
 mapping initial positions to current positions at time $t$. The
time interval $[t_{-},t_{+}]$ is part of the definition of the finite-time
dynamical system \eqref{eq:v(x,t)}. This interval may be a time scale
of interest or the maximum interval over which velocity data is available
from simulations or observations. 

The right Cauchy-Green strain tensor associated with the flow map
is defined as 
\begin{equation}
C_{t_{0}}^{t}(x_{0})=\left[\nabla F_{t_{0}}^{t}(x_{0})\right]^{T}\nabla F_{t_{0}}^{t}(x_{0}),\label{eq:CG}
\end{equation}
measuring Lagrangian strain in the velocity field. This tensor is
symmetric and positive definite \citep{Truesdell2004}\textbf{. }We
label the eigenvalues and eigenvectors of $C_{t_{0}}^{t}(\boldsymbol{x}_{0})$
as follows:
\[
C_{t_{0}}^{t}\xi_{i}=\lambda_{i}\xi_{i},\qquad0<\lambda_{1}\leq\lambda_{2},\quad i=1,2;
\]
\begin{equation}
\left|\xi_{i}\right|=1,\quad\xi_{2}=\Omega\xi_{1},\quad\Omega=\left(\begin{array}{rr}
0 & -1\\
1 & 0
\end{array}\right).\label{eq:C-G_invariants}
\end{equation}

\subsection{Elliptic LCSs\label{sub:Elliptic-LCSs}}
We seek positions of closed material lines at time $t_0$ that prevail
as coherent Lagrangian vortex boundaries (or \emph{elliptic LCSs})
over a time interval $[t_{0},t]\subset[t_{-},t_{+}]$. \citet{haller13:_coher_lagran} argue that such initial material line positions are closed stationary
curves of the averaged strain functional 
\[
Q(\gamma)=\frac{1}{\sigma}\int_{0}^{\sigma}\frac{\sqrt{\langle r'(s),C_{t_{0}}^{t}(r(s))r'(s)\rangle}}{\sqrt{\langle r'(s),r'(s)\rangle}}\text{d}s,
\]
obtained by averaging the tangential strain arising over $[t_{0},t]$
along closed material lines parametrized as $r(s)$ with $s\in[0,\sigma].$
Solutions to this variational problem turn out to be closed orbits
of one of two parametrized vector-field families 
\begin{equation}
\eta_{\pm}^{\lambda}=\sqrt{\frac{\lambda_{2}-\lambda^{2}}{\lambda_{2}-\lambda_{1}}}\xi_{1}\pm\sqrt{\frac{\lambda^{2}-\lambda_{1}}{\lambda_{2}-\lambda_{1}}}\xi_{2},\label{eq:eta}
\end{equation}
with $\lambda>0$ playing the role of a parameter. Such closed orbits
satisfy the differential equation 
\begin{equation}
r'=\eta_{\pm}^{\lambda}(r),\label{eq:etafields}
\end{equation}
which coincide with null-geodesics of the Lorentzian metric family
$e_{\lambda}(u,v)=\left\langle u,\left[D_{t_{0}}^{t}(r)-\lambda^{2}I\right]v\right\rangle .$
For this reason, we refer to the computation of elliptic LCSs as limit
cycles of \eqref{eq:etafields} as \emph{geodesic detection of elliptic
LCSs}.

Any orbit of \eqref{eq:etafields} turns out to stretch uniformly
under the flow map $F_{t_{0}}^{t}$. Specifically, any subset of an
orbit of \eqref{eq:etafields} increases its arclength precisely by
a factor of $\lambda$. For this reason, we refer to trajectories
of \eqref{eq:etafields} as $\lambda$-lines. Following \citep{haller13:_coher_lagran},
we call the outermost member of a closed family of $\lambda$-lines
a coherent Lagrangian vortex boundary.

\subsection{Hyperbolic LCSs\label{sub:Hyperbolic-LCSs}}
Next we consider positions of material lines at time $t_0$ that prevail
as most repelling or attracting material lines (or \emph{hyperbolic
LCSs}) over a time interval $[t_{0},t]\subset[t_{-},t_{+}]$. \citet{Farazmand2014} argue that hyperbolic LCSs are stationary
curves of the averaged shear functional 
\[
Q(\gamma)=\frac{1}{\sigma}\int_{0}^{\sigma}\frac{\langle r'(s),D_{t_{0}}^{t}(r(s))r'(s)\rangle}{\sqrt{\langle r'(s),C_{t_{0}}^{t}(r(s))r'(s)\rangle\langle r'(s),r'(s)\rangle}}\text{d}s,\qquad D_{t_{0}}^{t}=\frac{1}{2}[C_{t_{0}}^{t}\Omega-\Omega C_{t_{0}}^{t}],
\]
obtained by averaging the Lagrangian shear arising over $[t_{0},t]$
along closed material lines parametrized as $r(s)$ with $s\in[0,\sigma].$
More precisely, hyperbolic LCSs are stationary curves of $Q(\gamma)$
with respect to fixed-endpoint perturbations. We note that parabolic
LCSs (Lagrangian jet cores) are also stationary curves of $Q(\gamma)$,
but under variable endpoint perturbations (cf. \citep{Farazmand2014}). 

Solutions to this variational problem turn out to be orbits of the
$\xi_{1}$ or $\xi_{2}$ eigenvector field. Repelling LCSs (\emph{shrinklines})
are obtained as trajectories of the differential equation 
\begin{equation}
r'=\xi_{1}(r),\label{eq:strainline}
\end{equation}
and attracting LCSs (\emph{stretchlines}) are obtained as trajectories
of the differential equation
\begin{equation}
r'=\xi_{2}(r).\label{eq:stretchline}
\end{equation}

Shrinklines and stretchlines coincide with the null-geodesics of the
Lorentzian metric $h(u,v)=\left\langle u,D_{t_{0}}^{t}(r)v\right\rangle .$
For this reason, we refer to the computation of hyperbolic LCSs as
strongest normally-repelling or normally-attracting orbits of \eqref{eq:etafields}
as \emph{geodesic detection of hyperbolic LCSs}. 

To compute the normal repulsion of shrinklines, we note that an infinitesimal
normal perturbation to a shrinkline $\gamma$ at its point $r$ grows
under the flow map $F_{t_{0}}^{t}$ by a factor of $\lambda_{2}(r)$
in the direction normal to $F_{t_{0}}^{t}(\gamma).$ Similarly, small
normal perturbations to a stretchline decay by a factor $\lambda_{1}(r)$
in the direction normal to the evolving stretchline. 

\clearpage{}

\section{Numerical methods}
Here we describe a step-by-step numerical implementation of the geodesic
detection of elliptic and hyperbolic LCSs through the differential
equations \eqref{eq:etafields}, \eqref{eq:strainline}, and \eqref{eq:stretchline}.

\subsection{Computing the invariants of the Cauchy-Green strain tensor}
The common first step in calculating elliptic and hyperbolic LCSs
is the computation of the Cauchy-Green strain tensor field $C_{t_{0}}^{t}(x_{0})$,
as defined in \eqref{eq:CG}. The function performing this calculation
in LCS Tool is \inputencoding{latin9}\lstinline!eig_cgStrain!\inputencoding{utf8}.
The main steps executed by this function are enumerated in Table~\ref{t:Cauchy-Green algorithm},
while Table~\ref{t:eig_cgStrain syntax} summarizes the syntax of
\inputencoding{latin9}\lstinline!eig_cgStrain!\inputencoding{utf8}.

\begin{table}
\begin{enumerate}
\item Define a Cartesian grid for initial conditions of trajectories. Define
an auxiliary grid for differentiating with respect to initial conditions. 
\item Solve \eqref{eq:v(x,t)} starting from each grid point and auxiliary
grid point over the time interval $[t_{0},t${]}. This gives a discrete
approximation to the flow map $F_{t_{0}}^{t}(x_{0})$. 
\item Use finite differencing over the auxiliary grid to compute numerically
the derivative of the flow map $DF_{t_{0}}^{t}(x_{0})$. 
\item Compute the Cauchy-Green strain tensor field $C_{t_{0}}^{t}(x_{0})=\left(DF_{t_{0}}^{t}(x_{0})\right)^{T}DF_{t_{0}}^{t}(x_{0})$,
its eigenvalue field $\lambda_{1,2}(x_{0})$, and eigenvector fields
$\xi_{1,2}(x_{0})$ over the initial condition grid.
\end{enumerate}
\protect\caption{Algorithm to calculate the invariants of the Cauchy-Green strain tensor
field.}
\label{t:Cauchy-Green algorithm} 
\end{table}

\begin{table}
\begin{centering}
\begin{tabular}{|c|p{0.7\textwidth}|}
\hline 
\multicolumn{2}{|p{\textwidth}|}{\inputencoding{latin9}\lstinline![cgEigenvector,cgEigenvalue] = eig_cgStrain(derivative,domain,timespan,resolution)!\inputencoding{utf8}}\tabularnewline
\hline 
\inputencoding{latin9}\lstinline!derivative!\inputencoding{utf8}  & function handle for flow velocity equations\tabularnewline
\hline 
\inputencoding{latin9}\lstinline!domain!\inputencoding{utf8}  & $2\times2$ array to define flow domain\tabularnewline
\hline 
\inputencoding{latin9}\lstinline!timespan!\inputencoding{utf8}  & $1\times2$ array to define flow timespan\tabularnewline
\hline 
\inputencoding{latin9}\lstinline!resolution!\inputencoding{utf8}  & $1\times2$ array to define Cauchy-Green strain main grid resolution\tabularnewline
\hline 
\inputencoding{latin9}\lstinline!auxGridRelDelta!\inputencoding{utf8}  & optional scalar between 0 and 0.5 to specify auxiliary grid spacing.
Default is $10^{-2}$.\tabularnewline
\hline 
\inputencoding{latin9}\lstinline!eigenvalueFromMainGrid!\inputencoding{utf8}  & optional logical to control whether eigenvalues of Cauchy-Green strain
are calculated from main or auxiliary grid. Default is \inputencoding{latin9}\lstinline!true!\inputencoding{utf8}.\tabularnewline
\hline 
\inputencoding{latin9}\lstinline!incompressible!\inputencoding{utf8}  & optional logical to specify if incompressibility is imposed. Default
is \inputencoding{latin9}\lstinline!false!\inputencoding{utf8}.\tabularnewline
\hline 
\inputencoding{latin9}\lstinline!odeSolverOptions!\inputencoding{utf8}  & optional \inputencoding{latin9}\lstinline!odeset!\inputencoding{utf8}
structure to specify flow map integration parameters\tabularnewline
\hline 
\end{tabular}\protect\caption{Syntax of the function \lstinline!eigcgStrain!.}
\par\end{centering}
\centering{}\label{t:eig_cgStrain syntax} 
\end{table}

The Cartesian grid of initial conditions mentioned in Table~\ref{t:Cauchy-Green algorithm}
is rectangular, with user-defined vertical and horizontal ranges and
resolutions. The optimal resolution may be determined by a successive
doubling of the initial resolution until convergence of the extracted
LCSs is observed visually. If the domain of interest comprises only
a few expected LCSs, e.g., one vortex, then a resolution of about
500 grid point along the longest axis usually gives good results.
Otherwise, a higher resolution must be chosen.

\begin{figure}
\begin{centering}
\includegraphics[width=0.6\textwidth]{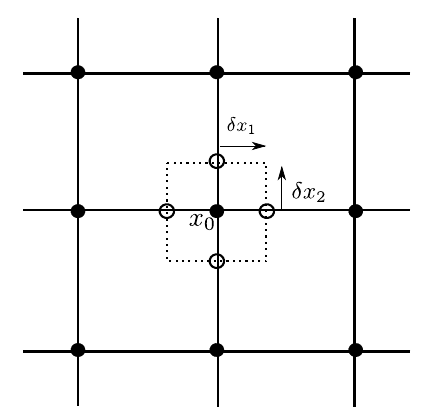} 
\par\end{centering}
\protect\caption{Illustration of the main grid (filled circles) and the auxiliary grid
(empty circles) used in the computation of the derivative of the flow
map in the definition \eqref{eq:CG} of the Cauchy--Green stain tensor.
The variable \lstinline!auxGridRelDelta! specifies the grid spacing
of the auxiliary grid relative to the main grid spacing.}
\label{f:main and auxiliary grids} 
\end{figure}

The auxiliary grid (cf. Table~\ref{t:Cauchy-Green algorithm}) comprises
four points placed symmetrically around each point of the Cartesian
grid (Figure~\ref{f:main and auxiliary grids}). These points are
used to achieve increased accuracy in the finite-difference approximation
\[
\nabla F_{t_{0}}^{t}(x_{0})\approx\left(\begin{array}{cc}
\frac{x_{1}(t;t_{0},x_{0}+\delta x_{1})-x_{1}(t;t_{0},x_{0}-\delta x_{1})}{2\vert\delta x_{1}\vert} & \frac{x_{1}(t;t_{0},x_{0}+\delta x_{2})-x_{1}(t;t_{0},x_{0}-\delta x_{2})}{2\vert\delta x_{2}\vert}\\
\frac{x_{2}(t;t_{0},x_{0}+\delta x_{1})-x^{2}(t;t_{0},x_{0}-\delta x_{1})}{2\vert\delta x_{1}\vert} & \frac{x_{2}(t;t_{0},x_{0}+\delta x_{2})-x_{2}(t;t_{0},x_{0}-\delta x_{2})}{2\vert\delta x_{2}\vert}
\end{array}\right)
\]
of $\nabla F_{t_{0}}^{t}$ at a point $x_{0}$ of the Cartesian grid.
Here $\delta x_{i}$ is a vector of length $\vert\delta x_{i}\vert>0$
that points from the Cartesian grid-point $x_{0}$ in the $i^{th}$
coordinate direction (Figure~\ref{f:main and auxiliary grids}).
Computational improvements arising from the use of the auxiliary grid
over simply using the nearest points of the main grid were reported
in Ref.\,\citep{farazmand12:_comput_lagran}. Experience suggests an auxiliary
grid spacing of 1-10\% of the main grid spacing.

The function \inputencoding{latin9}\lstinline!eig_cgStrain!\inputencoding{utf8}
of LCS Tool provides the option to calculate Cauchy--Green eigenvectors
from the auxiliary grid using eigenvalues calculated from the main
grid. We have found that for flows defined analytically, the eigenvalues
can reliably be calculated from the main grid. For the flows defined
by datasets, using the auxiliary grid for eigenvalue calculations
gives better results.

As mentioned above, a typical main grid for the Cauchy-Green strain
tensor comprises $500\times500$. This means that after the addition
of $4$ auxiliary grid point around each main grid points, eq. \eqref{eq:v(x,t)}
must be integrated over 1.25 million initial conditions. 

To avoid excessive computational times in MATLAB, we vectorize \eqref{eq:v(x,t)},
i.e., combine its right-hand-side evaluated over each initial points
into a single system of equations. The resulting system is composed
of independent blocks of two-dimensional first-order ODEs. We then
use MATLAB's \inputencoding{latin9}\lstinline!ode45!\inputencoding{utf8}
to perform trajectory integration from all grid points simultaneously.
This process typically takes five to ten minutes with default error
tolerances. A potential drawback of vector form integration is that
memory requirements may become excessive at high resolutions. Furthermore,
writing the velocity function in vector form is more error-prone than
the simpler two-dimensional form.

The Cauchy-Green strain tensor generically admits points singularities,
i.e., points where $C_{t_{0}}^{t}(x_{0})$ has repeated eigenvalues.
At these points the eigenvectors $\xi_{1}(x_{0})$ and $\xi_{2}(x_{0})$
are no longer well-defined. This generically arises at a finite set
of isolated points within the computational domain \citep{delmarcelle94},
and hence lie off the computational grid with probability one.

\subsubsection{Special case: incompressible velocity fields}
Incompressible flows (i.e. those satisfying $\nabla\cdot v=0$) admit
the relationship $\lambda_{1}(x_{0})\lambda_{2}(x_{0})=1$ at all
points of the computational domain \citep{arnold78:_mathem}. Incompressibility
can be computationally imposed by first calculating $\lambda_{2}(x_{0})$,
then setting $\lambda_{1}(x_{0})=1/\lambda_{2}(x_{0})$ and calculating
the strain eigenvectors $\xi_{2}$ from $\lambda_{2}$, then $\xi_{1}$
from the relationship in \eqref{eq:C-G_invariants}. Experience shows
that computing $\lambda_{i}$ in this order gives higher accuracy
than in the reverse order \citep{farazmand12:_comput_lagran}. 

At some grid points, $\lambda_{2}<1$ may occur due to numerical integration
errors. By setting the integration tolerances to smaller values, the
number of such grid points is reduced. Enforcing $\lambda_{2}\geq1$
everywhere, however, can incur excessive computational cost. To this
end, the function \inputencoding{latin9}\lstinline!eig_cgStrain!\inputencoding{utf8}
records the number of points with $\lambda_{2}<1$, providing a measure
for setting feasible integration tolerances.

\subsubsection{Special case: data-defined velocity fields}
Velocity fields defined by datasets require pre-processing before
they are used in the numerical integration of \eqref{eq:v(x,t)}.
This requires spatial and temporal interpolation that enables the
evaluation of the velocity function at arbitrary points in $U$ and
at arbitrary times between $t_{-}$ and $t_{+}$. In section \ref{sec:oceandataset},
we present an ocean dataset example with details of possible interpolation
functions.

\subsection{Computing elliptic LCSs}
As discussed in section \eqref{sub:Elliptic-LCSs}, positions of elliptic
LCSs at time $t_{0}$ are found as closed orbits of the $\eta_{\pm}^{\lambda}$
vector fields defined in \eqref{eq:eta}. We find such orbits by integrating
\eqref{eq:etafields} from points of an appropriately chosen section
(Poincare section), and evaluating the first return map (Poincar\'e
map) onto this section. A $\lambda$-line returning to its starting
point is then an elliptic LCS. The outermost member of a family of
closed $\lambda$-lines (obtained by varying $\lambda$) is a coherent
Lagrangian vortex boundary \citep{haller13:_coher_lagran}.

The main steps in calculating elliptic LCSs are enumerated in Table~\ref{t:Elliptic LCS algorithm}
and described in further detail below. The syntax of elliptic LCS
functions in LCS Tool is shown in Table~\ref{t:Elliptic LCS functions}.

\begin{table}
\begin{enumerate}
\item Position Poincare sections in flow domain to specify initial positions
of lambda-lines 
\item Integrate $\lambda$-lines tangent to $\eta_{\pm}^{\lambda}$ (see
~\eqref{eq:etafields}.) 
\item Calculate Poincare map 
\item Find closed orbits for fixed points of the Poincare map 
\item Identify outermost closed orbit on each Poincare section
\end{enumerate}
\protect\caption{Algorithm to calculate elliptic LCSs and coherent Lagrangian vortex
boundaries.}
\label{t:Elliptic LCS algorithm} 
\end{table}

\begin{table}
\begin{tabular}{|c|p{0.7\textwidth}|}
\hline 
\multicolumn{2}{|p{\textwidth}|}{\inputencoding{latin9}\lstinline![shearline.etaPos,shearline.etaNeg] = lambda_line(cgEigenvector,cgEigenvalue,lambda)!\inputencoding{utf8}}\tabularnewline
\hline 
\inputencoding{latin9}\lstinline!cgEigenvector!\inputencoding{utf8}  & array of Cauchy-Green strain eigenvectors\tabularnewline
\hline 
\inputencoding{latin9}\lstinline!cgEigenvalue!\inputencoding{utf8}  & array of Cauchy-Green strain eigenvalues\tabularnewline
\hline 
\inputencoding{latin9}\lstinline!lambda!\inputencoding{utf8}  & scalar lambda value in Equation~\eqref{eq:eta}\tabularnewline
\hline 
\hline 
\multicolumn{2}{|p{\textwidth}|}{\inputencoding{latin9}\lstinline![closedOrbits,orbits] = poincare_closed_orbit_multi(domain,resolution,shearline,PSList)!\inputencoding{utf8}}\tabularnewline
\hline 
\inputencoding{latin9}\lstinline!domain!\inputencoding{utf8}  & array to define flow domain\tabularnewline
\hline 
\inputencoding{latin9}\lstinline!resolution!\inputencoding{utf8}  & $1\times2$ array to define main grid resolution for Cauchy-Green
strain tensor\tabularnewline
\hline 
\inputencoding{latin9}\lstinline!shearline!\inputencoding{utf8}  & structure of arrays of $\eta_{+}^{\lambda}$ and $\eta_{-}^{\lambda}$
values on main grid\tabularnewline
\hline 
\inputencoding{latin9}\lstinline!PSList!\inputencoding{utf8}  & user-defined structure for Poincare section end-points, number of
$\lambda$-lines launched from Poincare section, and maximum closed$\lambda$-line
length\tabularnewline
\hline 
\inputencoding{latin9}\lstinline!nBisection!\inputencoding{utf8}  & optional number of bisection steps to refine zero crossings of Poincare
map. Default values: 5.\tabularnewline
\hline 
\inputencoding{latin9}\lstinline!dThresh!\inputencoding{utf8}  & optional threshold to discard discontinuous zero crossings of Poincare
map. Default value: $10^{-2}$.\tabularnewline
\hline 
\inputencoding{latin9}\lstinline!odeSolverOptions!\inputencoding{utf8}  & optional \inputencoding{latin9}\lstinline!odeset!\inputencoding{utf8}
structure to specify $\lambda$-line integration parameters\tabularnewline
\hline 
\inputencoding{latin9}\lstinline!periodicBc!\inputencoding{utf8}  & optional $1\times2$ logical array to specify periodic boundary conditions.
Default is \inputencoding{latin9}\lstinline![false,false]!\inputencoding{utf8}.\tabularnewline
\hline 
\end{tabular}\protect\caption{Syntax of LCS Tool's elliptic LCS functions.}
\label{t:Elliptic LCS functions} 
\end{table}

The first step is to define the position of Poincare sections in regions
where closed $\lambda$-lines are expected based on a visual analysis
of the orbit structure of the $\eta_{\pm}^{\lambda}$ vector field.
The Poincare section is to be oriented such that the first endpoint
is close to the center of the expected Lagrangian vortex, and the
second endpoint is outside this vortex. There is no automated procedure
implemented yet in LCS Tool for the positioning of Poincare sections
(see, however, \citep{Karrasch2014} for a recently developed algorithm).
Additionally, the number of lambda-lines launched from the Poincar\'e
section, \inputencoding{latin9}\lstinline!poincareSection.numPoints!\inputencoding{utf8},
must be defined. A reasonable default value is 100.

\begin{figure}
\begin{centering}
\includegraphics[width=0.8\textwidth]{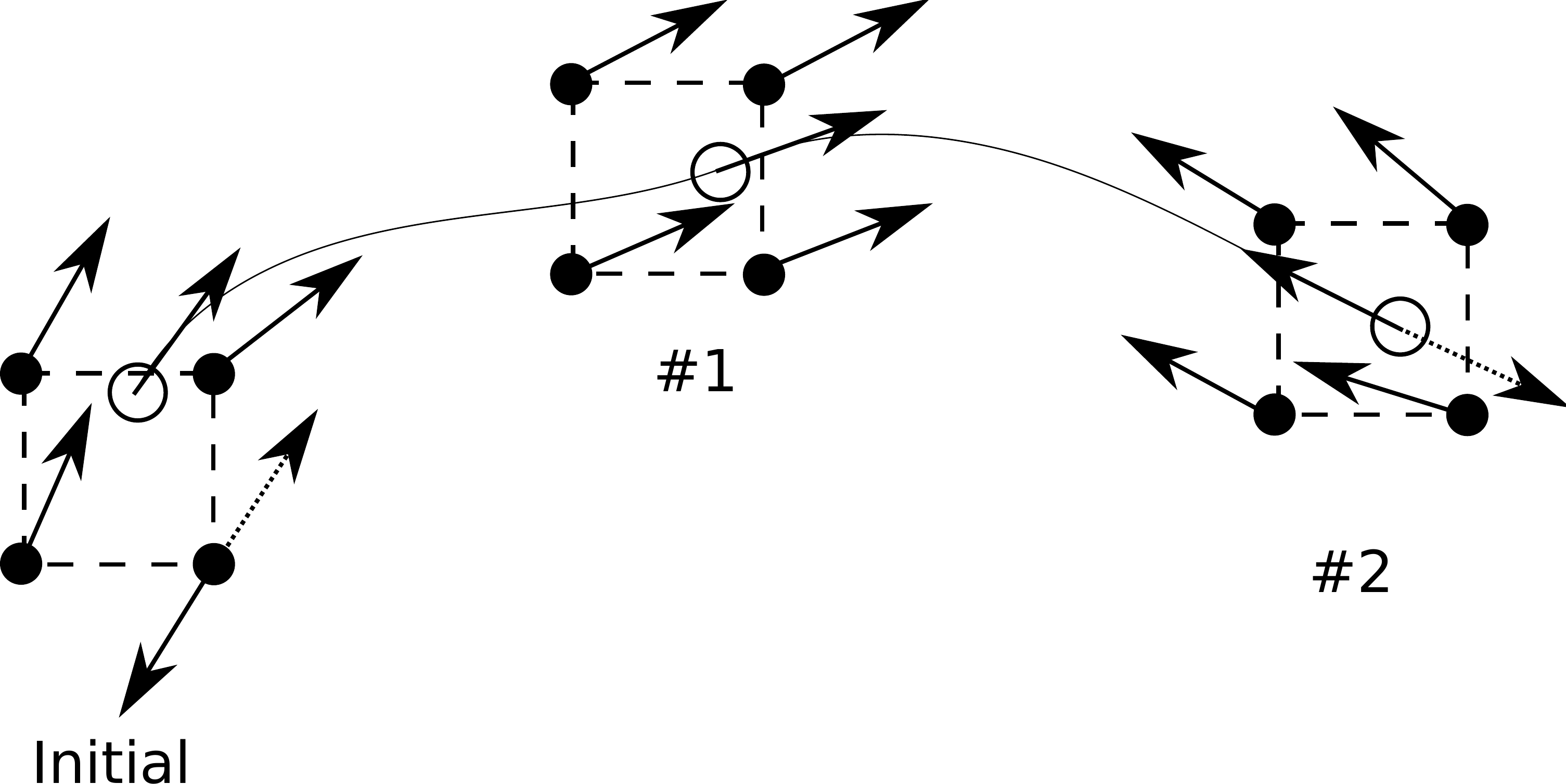} 
\par\end{centering}

\protect\caption{Schematic illustration of the variable-time-step $\lambda$-line integration.
At the initial point, there is an orientation discontinuity at the
lower-right grid point that must be corrected prior to linear interpolation.
At point \#1, no orientation discontinuities are present. At point
\#2 all interpolated $\eta_{\pm}^{\lambda}$ vectors must be rotated
by 180\degree\,to match the orientation of the trajectory.}
\label{f:variable step integration} 
\end{figure}

The second step is to integrate the $\lambda$-lines starting from
the Poincare section to obtain the corresponding Poincare map. Integration
of $\lambda$-lines is performed using the $\eta_{\pm}^{\lambda}$
vector fields defined in \eqref{eq:eta} over the main grid. The underlying
eigenvector fields $\xi_{i}(x_{0})$ have generic but removable orientation
discontinuities, which require monitoring and reorientation, as needed.
This process is sketched in Table~\ref{t:variable step integration}
and illustrated in Figure~\ref{f:variable step integration}. Linear
interpolation is used in the interpolation of $\eta_{\pm}^{\lambda}$
within a grid element, since using higher-order interpolation would
necessitate verifying that there are no orientation discontinuities
beyond the four nearest grid points. We identify orientation discontinuities
by checking the inner product of the $\eta_{\pm}^{\lambda}$ vectors
at adjacent grid points. Rotations exceeding 90\degree\, between
two such neighboring vectors are classified as orientation discontinuities
and are corrected before linear interpolation. When setting the Cauchy-Green
strain tensor main grid resolution, one may find it helpful to calculate
a histogram of eigenvector-field rotations to ensure that all rotations
are well below 90\degree.

\begin{table}
\begin{enumerate}
\item Linearly interpolate vector field orientation at initial position 
\item At next position, check whether vector field has rotated by over 90\degree,
if yes, flip the vector field orientation by 180\degree. 
\item Stop integration when $\lambda$-line returns to Poincare section,
$\lambda$-line reaches the domain boundary, or maximum integration
length has been reached. 
\end{enumerate}
\protect\caption{Algorithm used for variable time step integration of $\lambda$-lines.}
\label{t:variable step integration} 
\end{table}

\begin{figure}
\begin{centering}
\includegraphics[width=0.8\textwidth]{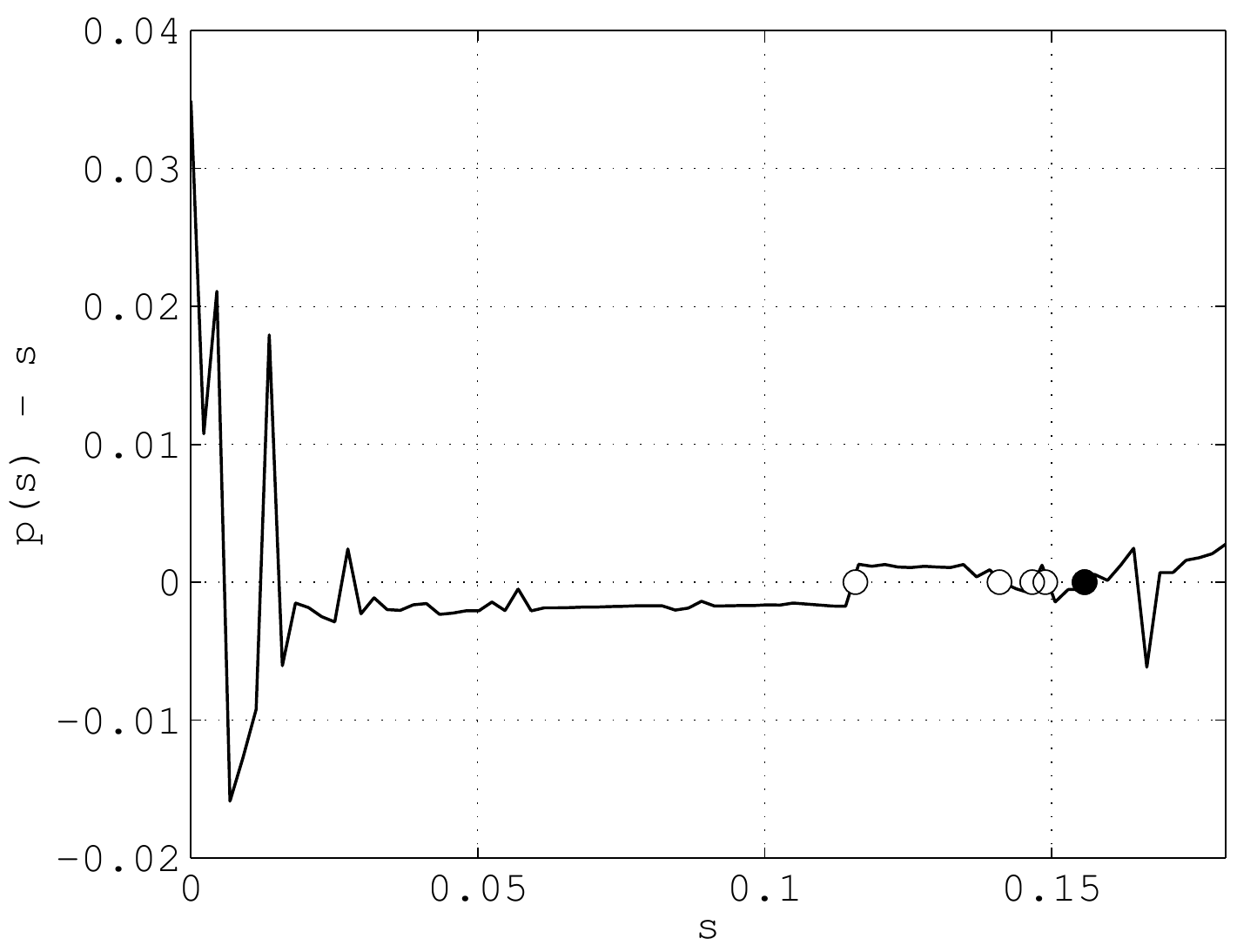} 
\par\end{centering}
\protect\caption{Example of a Poincare map obtained for an elliptic LCS region. Circle
markers indicate closed orbit positions. The filled circle indicates
the outermost fixed point of the Poincare map, marking the intersection
of a coherent Lagrangian cortex boundary with the Poincare section.}
\label{f:Poincare return map} 
\end{figure}

An example of a Poincare map produced from integration of the $\eta_{\pm}^{\lambda}$
field is shown in Figure~\ref{f:Poincare return map}. Most orbits
will return to the Poincare section and their integration will then
be stopped using the ODE event detection function of MATLAB. Some
orbits may, however, deviate far from the Poincare section and do
not return for any reasonable integration time. To control this behavior,
we specify a maximum orbit length, \inputencoding{latin9}\lstinline!poincareSection.orbitMaxLength!\inputencoding{utf8}
. In practice, viewing the Poincare section as the radius of a circle
and setting the maximum $\lambda$-line integration length to twice
the circumference gives good results. 

In Figure~\ref{f:Poincare return map}, circle markers indicate fixed
points of the Poincare map, i.e., points where the distance between
the final and the initial point of the orbit $P(s)-s$, is zero. The
function \inputencoding{latin9}\lstinline!poincare_closed_orbit_multi!\inputencoding{utf8}
performs the computations. As seen in Figure~\ref{f:Poincare return map},
LCS Tool uses a filtering parameter \inputencoding{latin9}\lstinline!dThresh!\inputencoding{utf8}
to discard sign changes of $P(s)-s$ that are likely due to numerical
sensitivity or a jump discontinuity of the Poincare map. Specifically,
the location of each detected zero crossing is first refined by the
bisection method. If after a predetermined number of iterations, \inputencoding{latin9}\lstinline!nBisection!\inputencoding{utf8}
(default value 5) the two points around the zero crossing still have
absolute values above \inputencoding{latin9}\lstinline!dThresh!\inputencoding{utf8},
the zero crossing is discarded. Once all valid closed $\lambda$-lines
have been located, the outermost closed $\lambda$-orbit associated
with every Poincare section is identified as a coherent Lagrangian
vortex boundary.

\subsection{Computing hyperbolic LCSs}
As discussed in section \eqref{sub:Hyperbolic-LCSs}, positions of
hyperbolic LCSs at time $t_{0}$ are found as the strongest repelling
orbits of the vector field (\ref{eq:strainline}) (repelling LCSs),
and strongest attracting orbits of the vector field \eqref{eq:stretchline}
(attracting LCSs). By repulsion and attraction we mean a property
of the LCS (as an evolving material line) under the flow map $F_{t_{0}}^{t}$.
We identify the strongest repelling shrinklines as those crossing
a local maximum of the $\lambda_{2}(x_{0})$ field. Similarly, we
identify the strongest attracting stretchlines as those crossing a
local minimum of the $\lambda_{1}(x_{0})$ field. These local maxima
and minima of the appropriate $\lambda_{i}(x_{0})$ eigenvalue field
can be thought of as the extensions of the concept of a saddle points
to the present finite-time, temporally aperiodic flow setting. 

The main steps of our hyperbolic LCS detection algorithm are enumerated
in Table~\ref{t:Hyperbolic LCS algorithm}. The basic function to
compute hyperbolic LCSs in LCS Tool is \inputencoding{latin9}\lstinline!seed_curves_from_lambda_max!\inputencoding{utf8},
with its syntax is given in Table~\ref{t:seed_curves_from_lambda_max syntax}.

\begin{table}
\begin{enumerate}
\item Define a local maximization distance. 
\item Find all points of the main grid that are local maxima of $\lambda_{2}$
within a circle whose radius is the local maximization distance. 
\item Define a maximum shrinkline length 
\item Integrate a shrinkline forward and backward according to Equation~\eqref{eq:strainline}
and using the largest $\lambda_{2}$ local maximum as the initial
position. Integrate until the shrinkline has attained the maximum
shrinkline length, or until it has reached the domain boundary. 
\item Flag any remaining local maxima of $\lambda_{2}$ within the maximization
distance of the shrinkline as ineligible initial positions for subsequent
shrinklines 
\item Continue integrating shrinklines using local maxima of $\lambda_{2}$
as initial positions until no eligible local maxima of $\lambda_{2}$
remain. 
\item Remove all shrinkline segments falling within elliptic LCSs. 
\end{enumerate}
\protect\caption{Algorithm to calculate initial positions of repelling LCSs at time
$t_{0}$. The algorithm for attracting LCSs is similar.}
\label{t:Hyperbolic LCS algorithm} 
\end{table}

\begin{table}
\begin{centering}
\begin{tabular}{|c|p{0.7\textwidth}|}
\hline 
\multicolumn{2}{|p{\textwidth}|}{\inputencoding{latin9}\lstinline![curvePosition,curveInitialPosition] = seed_curves_from_lambda_max(distance,cgEigenvalue,cgEigenvector,flowDomain,flowResolution)!\inputencoding{utf8}}\tabularnewline
\hline 
\inputencoding{latin9}\lstinline!distance!\inputencoding{utf8}  & threshold distance for placement of $\lambda_{2}(x_{0})$ maxima\tabularnewline
\hline 
\inputencoding{latin9}\lstinline!cgEigenvalue!\inputencoding{utf8}  & array of Cauchy-Green strain eigenvalues\tabularnewline
\hline 
\inputencoding{latin9}\lstinline!cgEigenvector!\inputencoding{utf8}  & array of Cauchy-Green strain eigenvectors\tabularnewline
\hline 
\inputencoding{latin9}\lstinline!flowDomain!\inputencoding{utf8}  & $2\times2$ array to define flow domain\tabularnewline
\hline 
\inputencoding{latin9}\lstinline!flowResolution!\inputencoding{utf8}  & $1\times2$ array to define Cauchy-Green strain main grid resolution\tabularnewline
\hline 
\inputencoding{latin9}\lstinline!periodicBc!\inputencoding{utf8}  & optional $1\times2$ logical array to specify periodic boundary conditions.
Default is \inputencoding{latin9}\lstinline![false,false]!\inputencoding{utf8}.\tabularnewline
\hline 
\inputencoding{latin9}\lstinline!nMaxCurves!\inputencoding{utf8}  & optional maximum number of curves (i.e. shrinklines of stretchlines)
to generate. Default is \inputencoding{latin9}\lstinline!numel(cgEigenvalue)!\inputencoding{utf8}.\tabularnewline
\hline 
\inputencoding{latin9}\lstinline!odeSolverOptions!\inputencoding{utf8}  & optional \inputencoding{latin9}\lstinline!odeset!\inputencoding{utf8}
structure to specify flow map integration parameters\tabularnewline
\hline 
\end{tabular}
\par\end{centering}
\protect\caption{Syntax of the function \lstinline!seed_curves_from_lambda_max!}
\label{t:seed_curves_from_lambda_max syntax} 
\end{table}

\begin{figure}[hbt]
\centering 
\includegraphics[width=1.0\textwidth]{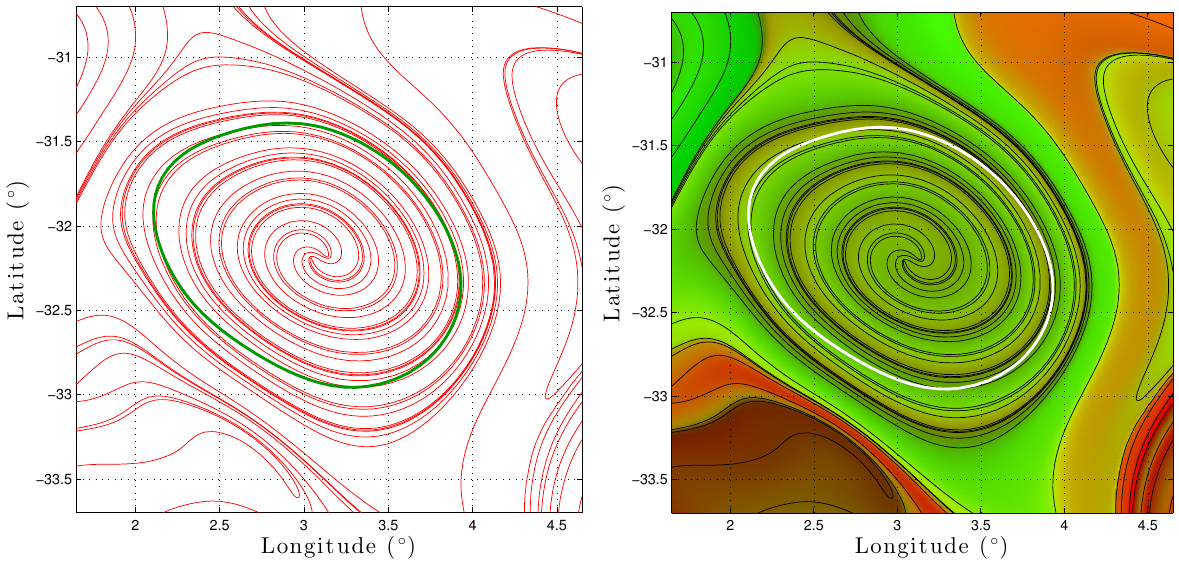}
\protect\caption{(Left) Elliptic LCS (green) with repelling LCSs (red) continuing into
its interior. (Right) Final positions $x(t)=\left(x_{1}(t),x_{2}(t)\right)=F_{t_{0}}^{t}(x_{0})$
of tracer particles, with their $x_{1}(t)$-position encoded in red
and $x_{2}(t)$-position encoded in green color. Note the continued
impact of the repelling hyperbolic LCSs (black) on tracer patterns inside the elliptic
LCS (white). }
\label{fig:ocean_dataset_colortracer} 
\end{figure}

In classical, infinite-time dynamical systems, elliptic and hyperbolic
invariant manifolds cannot intersect, as the trajectories in their
intersection would then have to follow two different asymptotic behaviors
simultaneously. In finite-time dynamical systems, hyperbolic LCSs
may well intersect elliptic LCSs, with hyperbolic LCSs continuing
to act as cores of mixing patterns observed in the interior of elliptic
LCSs. Motivated by convention and for clarity in visualization, however,
the default setting of LCS Tool removes hyperbolic LCS segments from
the interior of elliptic LCSs. These removed segments can be turned
back on demand, as shown in Figure~\ref{fig:ocean_dataset_colortracer}. 

\clearpage{}

\section{Examples}
In this section we present examples of the use of LCS Tool to obtain
LCSs in three different examples: a double gyre, a jet, and an oceanic
geostrophic flow. All three examples are available as demonstration
files in the demo folder of LCS Tool. By executing these scripts,
readers can follow LCS Tool computations step-by-step. Additionally,
LCS Tool includes scripts to animate the three flows.

\subsection{Double gyre}
The double gyre is a model for a time-dependent two-gyre system observed
in geophysical flows \citep{shadden05:_defin_lagran_lyapun}. The
model consists of two counter-rotating sinusoidal vortices with a
harmonically oscillating line in-between. Lagrangian particle motions
satisfy the non-autonomous dynamical system 
\begin{equation}
\begin{split}\dot{x}=-\pi A\sin[\pi f(x,t)]\cos(\pi y),\\
\dot{y}=\pi A\cos[\pi f(x,t)]\sin(\pi y)\frac{\partial f(x,t)}{\partial x},\\
f(x,t)=\epsilon\sin(\omega t)x^{2}+[1-2\epsilon\sin(\omega t)]x.
\end{split}
\label{eq:double gyre derivative equations}
\end{equation}
The MATLAB function describing this model velocity field is given
in Listing~\ref{l:double gyre derivative}, specifying the right-hand-side
of the particle ODE in a way that supports vectorized integration.

\begin{lstlisting}[breaklines=true,caption={Definition of double gyre flow.},captionpos=b,label={l:double gyre derivative},language=Matlab]
function derivative_ = derivative(t,x,epsilon,amplitude,omega)

idx1 = 1:2:numel(x)-1;
idx2 = 2:2:numel(x);

a = epsilon*sin(omega*t);
b = 1 - 2*epsilon*sin(omega*t);
forcing = a*x(idx1).^2 + b*x(idx1);

derivative_ = nan(size(x));

derivative_(idx1) = -pi*amplitude*sin(pi*forcing).*cos(pi*x(idx2));
derivative_(idx2) = pi*amplitude*cos(pi*forcing).*sin(pi*x(idx2)).*(2*a*x(idx1) + b);
\end{lstlisting}

In what follows, the parameter values are: $A=0.1$, $\epsilon=0.1$,
$\omega=\pi/5$. The flow timespan is $t\in[0,10]$ and the domain
is $x\in[0,2]$, $y\in[0,1]$. By examining the FTLE field (which
is optionally provided by LCS Tool), we position Poincare sections
to capture elliptic LCSs. An LCS Tool demo script to perform this
operation is given in Listing~\ref{l:double gyre lambda-line LCS}
where two Poincare sections are defined (ll.\,18-19). The free stretching
parameter $\lambda$ (Eq. \ref{eq:etafields}) is varied over the
range $[0.93,1.07]$ with increments of $0.01$ (ll.\,27-28). In
a loop (ll.\,35-39), closed orbits for all predefined $\lambda$
values are computed and the outermost closed orbit is kept as the
Lagrangian vortex boundary. In this script, we have chosen small values
for the error tolerances of the integration of the Cauchy-Green strain
tensor (l.\,14) and $\lambda$-lines (l.\,26) to facilitate convergence.

\begin{lstlisting}[breaklines=true,caption={Double gyre script for elliptic $\lambda$-line LCSs},captionpos=b,label={l:double gyre lambda-line LCS},language=Matlab]
%% Input parameters
epsilon = .1;
amplitude = .1;
omega = pi/5;
domain = [0,2;0,1];
resolution = [500,250];
timespan = [0,10];

%% Velocity definition
lDerivative = @(t,x,~)derivative(t,x,false,epsilon,amplitude,omega);
incompressible = true;

%% LCS parameters
cgStrainOdeSolverOptions = odeset('relTol',1e-5);

% Lambda-lines
poincareSection = struct('endPosition',{},'numPoints',{},'orbitMaxLength',{});
poincareSection(1).endPosition = [0.55,0.55;0.1,0.1];
poincareSection(2).endPosition = [1.53,.45;1.95,0.05];
[poincareSection.numPoints] = deal(100);
nPoincareSection = numel(poincareSection);
for i = 1:nPoincareSection
    rOrbit = hypot(diff(poincareSection(i).endPosition(:,1)),diff(poincareSection(i).endPosition(:,2)));
    poincareSection(i).orbitMaxLength = 2*(2*pi*rOrbit);
end
lambdaLineOdeSolverOptions = odeset('relTol',1e-6);
lambdaStep = 0.01;
lambdaRange = 0.93:lambdaStep:1.07;

%% Cauchy-Green strain eigenvalues and eigenvectors
[cgEigenvector,cgEigenvalue] = eig_cgStrain(lDerivative,domain,resolution,timespan,'incompressible',incompressible,'odeSolverOptions',cgStrainOdeSolverOptions);

%% Lambda-line LCSs

for lambda = lambdaRange
...        
    closedLambdaLineCandidate = poincare_closed_orbit_multi(domain,resolution,shearline,poincareSection,'odeSolverOptions',lambdaLineOdeSolverOptions,'showGraph',showGraph);
...
end
\end{lstlisting}

\begin{figure}[hbt]
\centering 
\includegraphics[width=1\textwidth]{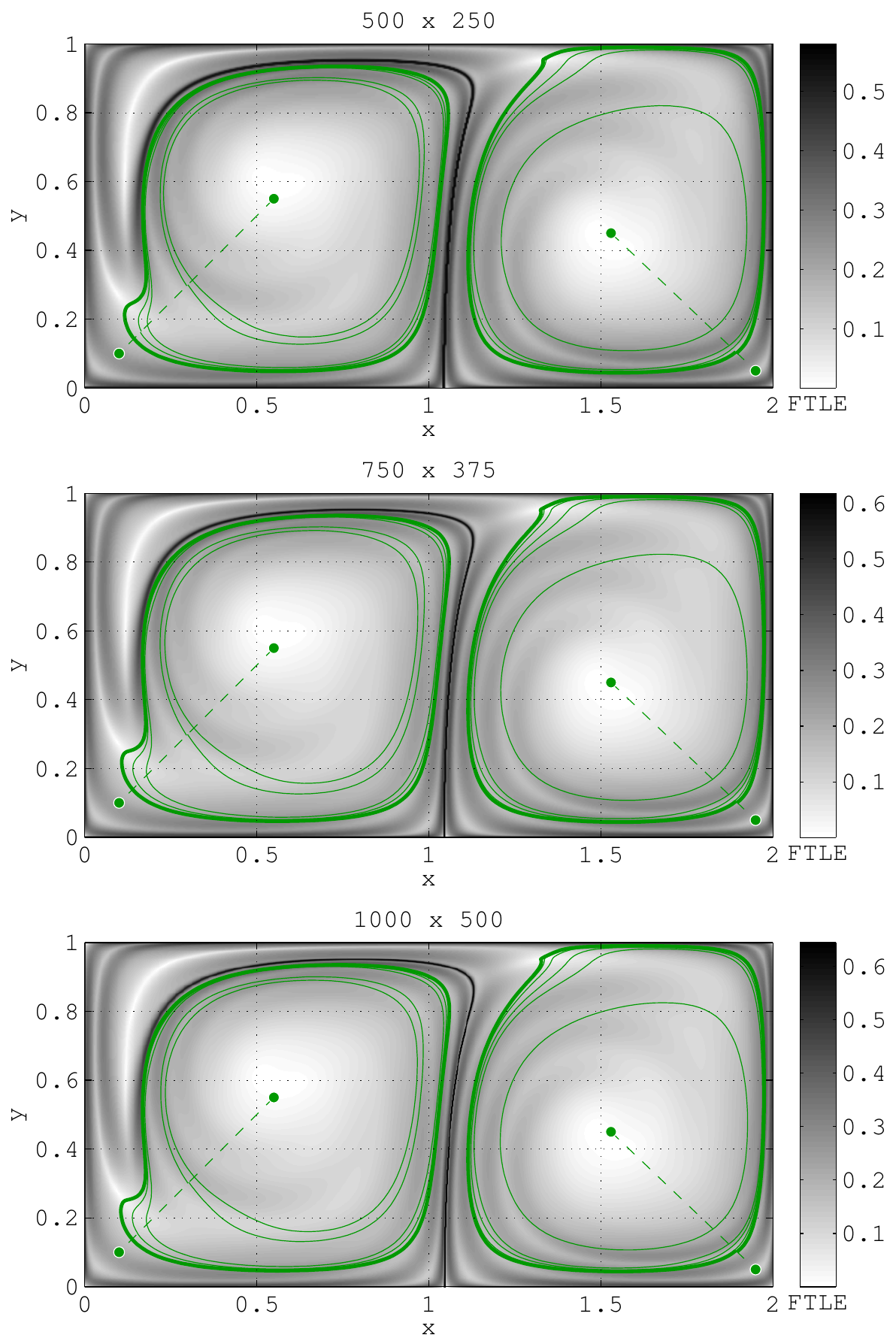}
\protect\caption{Convergence of closed $\lambda$-lines for increasing main-grid resolution
in the double-gyre model. The outermost closed $\lambda$-line (bold
green line) is the vortex boundary. (Top) {[}500x250{]}, $\lambda$-values
of closed orbits from interior to exterior, left gyre: $\lambda=(0.97, 0.98, 0.99, 1.00, 1.01)$,
right gyre: $\lambda=(1.00, 1.01, 1.02, 1.03, 1.04)$; (Middle) {[}750x375{]},
left: $\lambda=(0.97, 0.98, 0.99, 1.00, 1.01)$, right: $\lambda=(1.00, 1.01, 1.02, 1.03, 1.04, 1.05)$;
(Bottom) {[}1000x500{]}, left: $\lambda=(0.97, 0.98, 0.99, 1.00, 1.01)$,
right: $\lambda=(1.00, 1.01, 1.02, 1.03, 1.04, 1.05)$.}
\label{fig:double_gyre_lambda_lcs_convergence} 
\end{figure}

In Figure~\ref{fig:double_gyre_lambda_lcs_convergence}, the resolution
of the Cauchy-Green strain tensor is varied from $[500\times250]$
to $[1000\times500]$. The location of the outermost closed $\lambda$-line
changes insignificantly, demonstrating convergence. For all tested
resolutions the $\lambda$ values of the outermost closed orbits are
constant. This is also suggests that the lowest tested resolution
of $[500\times250]$ is sufficient to identify the elliptic LCS in
this example.

\begin{figure}
\centering
\includegraphics[width=1\textwidth]{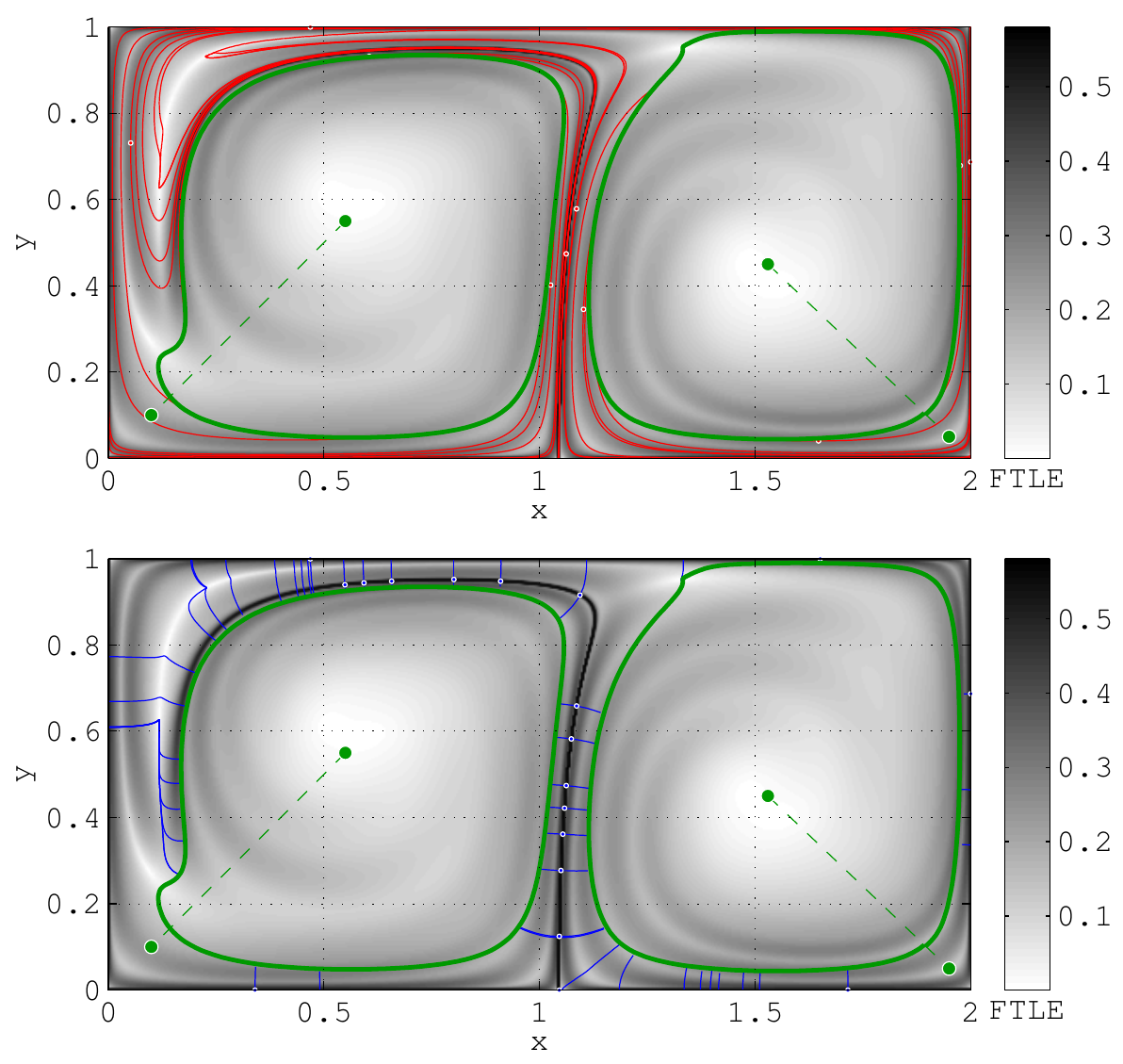}
\protect\caption{LCSs in the double gyre flow. Resolution $[500\times250]$. $\lambda = 1.00$
and $1.04$ for the left and the right gyre. $\lambda \in [0.93,1.07],\,\Delta\lambda=0.01$.
(Top) Elliptic (green) and repelling (red) LCSs with the FTLE shown
in the background. (Bottom) Elliptic and attracting (blue) LCSs.}
\label{fig:double_gyre_lambda_strain_lcs} 
\end{figure}

Figure \ref{fig:double_gyre_lambda_strain_lcs} shows the complete
picture of elliptic and hyperbolic LCSs of the double gyre at this
resolution. Listing~\ref{lst:double_gyre_lambda_strain_stretch_lcs}\textbf{
}lists the corresponding LCS Tool commands. The maximal length of
shrinklines and stretchlines, \lstinline!strainlineMaxLength! and
\lstinline!stretchlineMaxLength!, is set to $20$, a multiple of
the domain size, since the hyperbolic LCSs may wind around vortices
several times. 

As usual, the local maximization distance is set larger for stretchlines
than for shrinklines (cf. l.\,4 and l\,9). The purpose of the maximization
distance is to obtain spatially separated LCSs and to avoid a dense
tangle of lines that basically indicates the same hyperbolic LCS (cf.
Table\,\ref{t:Hyperbolic LCS algorithm}, point 5). Setting the local
maximization distance for stretchlines larger than for shrinklines
allows obtaining a comparable number of stretchlines and shrinklines
overall in the flow domain (recall that hyperbolic LCS seed points
are discarded if they are within the local maximization distance of
an existing hyperbolic LCS.) Shrinklines are locally tangent to ridges
of $\lambda_{2}$ maxima, whereas stretchlines are locally normal
to these ridges. Setting the local maximization distance of stretchlines
and shrinklines equal would therefore produce a greater number of
stretchlines than shrinklines overall.

\begin{lstlisting}[caption={Excerpt from demo script to compute hyperbolic LCS for the double gyre flow.},label={lst:double_gyre_lambda_strain_stretch_lcs}]
% Shrinklines
shrinklineMaxLength = 20;
gridSpace = diff(domain(1,:))/(double(resolution(1))-1);
shrinklineLocalMaxDistance = 2*gridSpace;
strainlineOdeSolverOptions = odeset('relTol',1e-6);

% Stretchlines
stretchlineMaxLength = 20;
stretchlineLocalMaxDistance = 10*gridSpace;
stretchlineOdeSolverOptions = odeset('relTol',1e-6);

%% Hyperbolic shrinkline LCSs
[shrinklineLcs,shrinklineLcsInitialPosition] = seed_curves_from_lambda_max(shrinklineLocalMaxDistance,shrinklineMaxLength,cgEigenvalue(:,2),cgEigenvector(:,1:2),domain,resolution,'odeSolverOptions',shrinklineOdeSolverOptions);
...
%% Hyperbolic stretchline LCSs
[stretchlineLcs,stretchlineLcsInitialPosition] = seed_curves_from_lambda_max(stretchlineLocalMaxDistance,stretchlineMaxLength,-cgEigenvalue(:,1),cgEigenvector(:,3:4),domain,resolution,'odeSolverOptions',stretchlineOdeSolverOptions);
...
\end{lstlisting}

\clearpage{}

\subsection{Bickley jet}
The Bickley jet models a meandering zonal jet flanked above and below
by counter-rotating vortices. This is an idealized model of geophysical
flows such as the Gulf Stream or the polar night jet perturbed by
a Rossby wave \citep{Castillo-Negrete1993,beron-vera10:_invar_lagran}.

The velocity is given by $v(x,y,t)=(-\partial_{y}\psi,\partial_{x}\psi)$
where 
\begin{gather*}
\psi(x,y,t)=\psi_{0}(x,y)+\psi_{1}(x,y,t),\\
\psi_{0}(x,y)=c_{3}y-UL_{y}\tanh\frac{y}{L_{y}}+\epsilon_{3}UL_{y}\mathrm{sech}^{2}\frac{y}{L_{y}}\cos k_{3}x,\\
\psi_{1}(x,y,t)=UL_{y}\mathrm{sech}^{2}\frac{y}{L_{y}}\mathrm{Re}\left[\sum_{n=1}^{2}\epsilon_{n}f_{n}(t)e^{ik_{n}x}\right].
\end{gather*}
As a forcing function, we choose a solution running on the chaotic
attractor of the damped-forced Duffing oscillator. Specifically, we
let 
\begin{gather*}
\frac{d\phi_{1}}{dt}=\phi_{2},\\
\frac{d\phi_{1}}{dt}=-0.1\phi_{2}-\phi_{1}^{3}+11\cos(t),\\
f_{1,2}(t)=2.625\times10^{-2}\phi_{1}(t/6.238\times10^{5}).
\end{gather*}
The parameter values we use are: $U=62.66$, $c_{2}=0.205U$, $c_{3}=0.461U$,
$L_{y}=1.77\times10^{6}$, $\epsilon_{1}=0.0075$, $\epsilon_{2}=0.04$,
$\epsilon_{3}=0.3$, $L_{x}=6.371\times10^{6}\pi$, $k_{n}=2n\pi/L_{x}$,
$\sigma_{1}=0.5k_{2}(c_{2}-c_{3})$, $\sigma_{2}=2\sigma_{1}$.

The integration time is $T=4L_x/U$, a multiple of a the eddy turnover
time in the flow (see also LCS Tool file \lstinline{flow_animation.m}).
Listing \ref{lst:BickleyJet} shows the code of LCS Tool in which
the chaotically perturbed velocity field is defined (ll.\,11-18),
the boundaries are set to periodic in the x-direction (l.\,20), and
five Poincare sections are defined where we expect coherent vortices
(ll.\,46-50). $\lambda$-values for the closed orbit detection are
varied over the range $[0.80,1.20]$ with a step of $0.01$.

\begin{figure}
\centering
\includegraphics[width=1\textwidth]{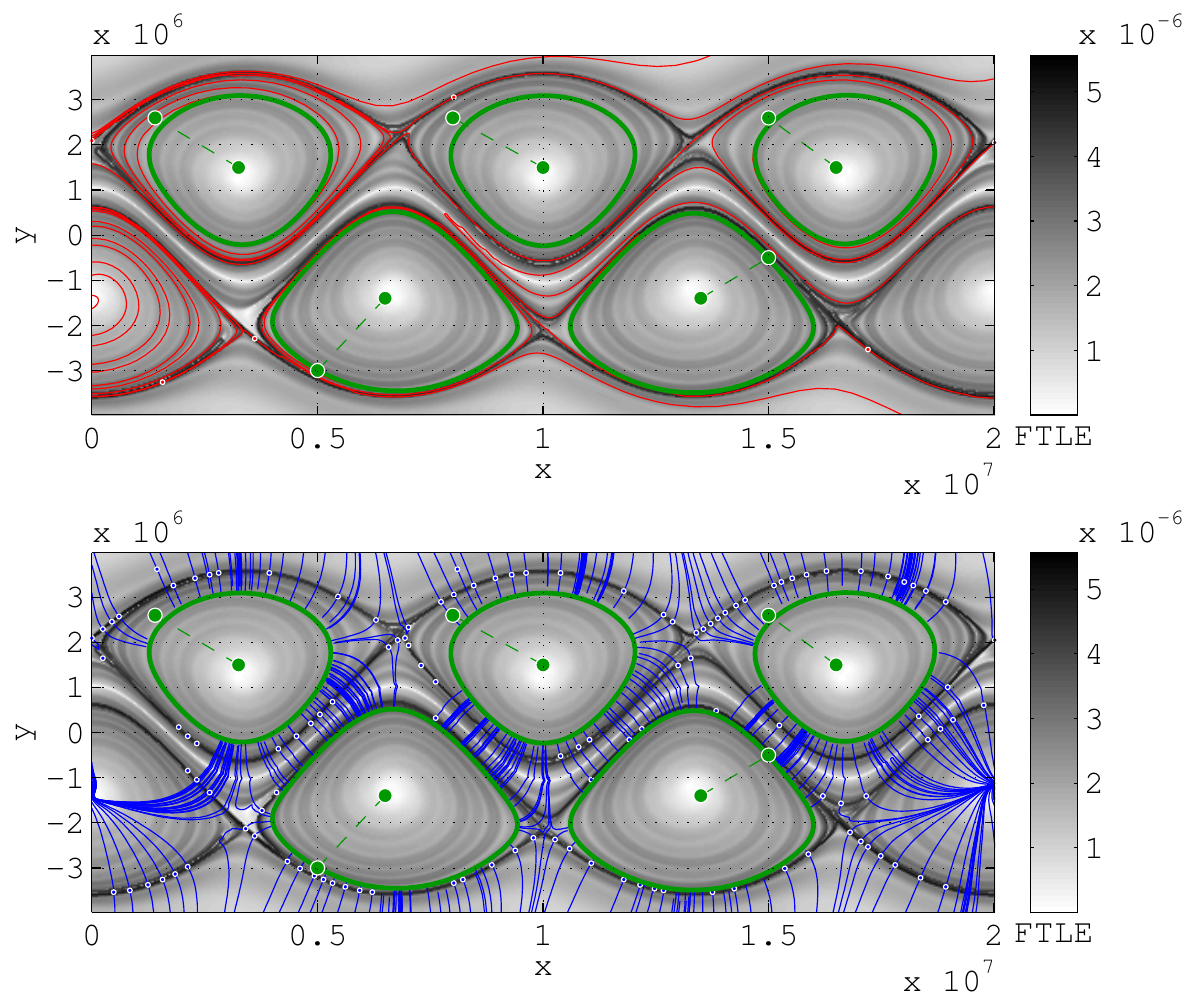}
\protect\caption{Elliptic and hyperbolic (top: red - repelling, bottom: blue - attracting)
LCSs in the Bickley jet. $\lambda$-valuestrains of coherent eddies
from left to right [0.95,\,0.80,\,0.94,\,0.80,\,0.94], $\lambda \in [0.80,1.20],\,\Delta\lambda=0.01$.}
\label{fig:Bickley_jet_lambda_LCS_full} 
\end{figure}

Figure~\ref{fig:Bickley_jet_lambda_LCS_full} shows elliptic and
hyperbolic LCSs of the Bickley jet with the FTLE field in the background.

\begin{lstlisting}[caption={Bickley jet script for hyperbolic and elliptic LCSs.},label={lst:BickleyJet}]
%% Input parameters
u = 62.66;
lengthX = pi*earthRadius;
lengthY = 1.77e6;
epsilon = [.075,.4,.3];
domain = [0,lengthX;[-1,1]*2.25*lengthY];
resolutionX = 500;
timespan = [0,4*lengthX/u];

%% Velocity definition
perturbationCase = 3;
phiTimespan = [0,25];
phiInitial = [0,0];
phiSol = ode45(@d_phi,phiTimespan,phiInitial);
timeResolution = 1e5;
phi1 = deval(phiSol,linspace(phiTimespan(1),phiTimespan(2),timeResolution),1);
phi1Max = max(phi1);
lDerivative = @(t,x,~)derivative(t,x,false,u,lengthX,lengthY,epsilon,perturbationCase,phiSol,phi1Max);
incompressible = true;
periodicBc = [true,false];

%% Cauchy-Green strain eigenvalues and eigenvectors
[cgEigenvector,cgEigenvalue] = eig_cgStrain(lDerivative,domain,resolution,timespan,'incompressible',incompressible);

%% LCS parameters
% Lambda-lines
lambdaStep = 0.01;
lambdaRange = 0.80:lambdaStep:1.10;
lambdaLineOdeSolverOptions = odeset('relTol',1e-6);

% Shrinklines
shrinklineMaxLength = 1e8;
shrinklineLocalMaxDistance = 8*gridSpace;
shrinklineOdeSolverOptions = odeset('relTol',1e-4);

% Stretchlines
stretchlineMaxLength = 1e8;
stretchlineLocalMaxDistance = 4*gridSpace;
stretchlineOdeSolverOptions = odeset('relTol',1e-4);

%% Lambda-line LCSs
% Define Poincare sections; first point in center of elliptic region and
% second point outside elliptic region
poincareSection = struct('endPosition',{},'numPoints',{},'orbitMaxLength',{});

poincareSection(1).endPosition = [3.25e6, 1.5e6;  1.4e6,  2.6e6];
poincareSection(2).endPosition = [6.50e6,-1.4e6;  5.0e6, -3.0e6];
poincareSection(3).endPosition = [1.0e7,  1.5e6;  8.0e6,  2.6e6];
poincareSection(4).endPosition = [1.35e7,-1.4e6;  1.5e7, -0.5e6];
poincareSection(5).endPosition = [1.65e7, 1.5e6;  1.5e7,  2.6e6];

% Number of orbit seed points along each Poincare section
[poincareSection.numPoints] = deal(80);

% Set maximum orbit length to twice the expected circumference
nPoincareSection = numel(poincareSection);
for i = 1:nPoincareSection
    rOrbit = hypot(diff(poincareSection(i).endPosition(:,1)),diff(poincareSection(i).endPosition(:,2)));
    poincareSection(i).orbitMaxLength = 2*(2*pi*rOrbit);
end

...
for lambda = lambdaRange    
...    
    [shearline.etaPos,shearline.etaNeg] = lambda_line(cgEigenvector,cgEigenvalue,lambda);          
    closedLambdaLineCandidate = poincare_closed_orbit_multi(domain,resolution,shearline,poincareSection,'odeSolverOptions',lambdaLineOdeSolverOptions);        
...    
end
...   
%% Hyperbolic shrinkline LCSs
[shrinklineLcs,shrinklineLcsInitialPosition] = seed_curves_from_lambda_max(shrinklineLocalMaxDistance,shrinklineMaxLength,cgEigenvalue(:,2),cgEigenvector(:,1:2),domain,resolution,'periodicBc',periodicBc,'odeSolverOptions',shrinklineOdeSolverOptions);

%% Hyperbolic stretchline LCSs
[stretchlineLcs,stretchlineLcsInitialPosition] = seed_curves_from_lambda_max(stretchlineLocalMaxDistance,stretchlineMaxLength,-cgEigenvalue(:,1),cgEigenvector(:,3:4),domain,resolution,'periodicBc',periodicBc,'odeSolverOptions',stretchlineOdeSolverOptions);
\end{lstlisting}

\clearpage{}

\subsection{Ocean velocity data from satellite altimetry}
\label{sec:oceandataset}
This last example demonstrates the use of LCS Tool on velocity data
derived from satellite-observed sea-surface heights under the geostrophic
approximation. In contrast to the previous two analytic examples,
the velocity field is available only with discrete temporal and spatial
resolution. Our region of interest is a small domain in the South
Atlantic Ocean, where exceptionally coherent eddies (Agulhas rings)
were recently found by \citep{haller13:_coher_lagran} using the theory
we surveyed in Section \ref{sub:Elliptic-LCSs}.

In the geostrophic approximation, the sea-surface height $\eta$ serves
as a stream-function for surface velocities. In a longitude-latitude
$(\varphi,\theta)$ coordinate system, the evolution of a fluid particle
is given by 
\begin{eqnarray}
\dot{\varphi}(\varphi,\theta,t)=-\frac{g}{R^{2}f(\theta)\cos\theta}\partial_{\theta}\eta(\varphi,\theta,t)\label{eq:dtlon}\\
\dot{\theta}(\varphi,\theta,t)=+\frac{g}{R^{2}f(\theta)\cos\theta}\partial_{\varphi}\eta(\varphi,\theta,t)\label{eq:dtlat}
\end{eqnarray}
where $g$ is the constant of gravity, $R$ is the mean radius of
the Earth, and $f(\theta)\equiv2\Omega\sin\theta$ is the Coriolis
parameter, with $\Omega$ denoting the Earth's mean angular velocity.

The data is given at a spatial resolution of $1/4\degree$ and a temporal
resolution of $7\,\mathrm{days}$. Due to the discrete data, defining
the right-hand side of Equation~\eqref{eq:dtlon} and~\eqref{eq:dtlat}
involves spline interpolation in space and time. An interpolant is
generated first, then the function \inputencoding{latin9}\lstinline!flowdata_derivative!\inputencoding{utf8}
evaluates the interpolants for the zonal and meridional velocity at
the needed coordinates. Listing \ref{l:griddedInterp}\textbf{ }shows
the relevant part of the code in LCS Tool's ocean demo file \inputencoding{latin9}\lstinline!hyperbolic_shear_lcs.m!\inputencoding{utf8}.
In lines 8-11, the commands for the interpolation of the velocity
data set are given.

\begin{lstlisting}[caption={LCS Tool demo script to compute LCS from an ocean data set.},label={l:griddedInterp}]
% Input parameters
domain = [0,6;-34,-28];
resolution = [400,400];
timespan = [100,130];

% Velocity definition
load('ocean_geostrophic_velocity.mat');
...
interpMethod = 'spline';
vlon_interpolant = griddedInterpolant({time,lat,lon},vlon,interpMethod);
vlat_interpolant = griddedInterpolant({time,lat,lon},vlat,interpMethod);
lDerivative = @(t,x,~)flowdata_derivative(t,x,vlon_interpolant,vlat_interpolant);
incompressible = true;

% LCS parameters
% Cauchy-Green strain
cgEigenvalueFromMainGrid = false;
cgAuxGridRelDelta = 0.01;

% Lambda-lines
lambdaStep = 0.02; lambdaRange = 0.90:lambdaStep:1.10;
lambdaLineOdeSolverOptions = odeset('relTol',1e-6);

% Shrinklines
shrinklineMaxLength = 20;
gridSpace = diff(domain(1,:))/(double(resolution(1))-1);
shrinklineLocalMaxDistance = 2*gridSpace;
shrinklineOdeSolverOptions = odeset('relTol',1e-4);

% Stretchlines
stretchlineMaxLength = 20;
stretchlineLocalMaxDistance = 4*gridSpace;
stretchlineOdeSolverOptions = odeset('relTol',1e-4);
...
% Cauchy-Green strain eigenvalues and eigenvectors
[cgEigenvector,cgEigenvalue] = eig_cgStrain(lDerivative,domain,resolution,timespan,'incompressible',incompressible,'eigenvalueFromMainGrid',cgEigenvalueFromMainGrid,'auxGridRelDelta',cgAuxGridRelDelta);

% Lambda-line LCSs
% Define Poincare sections; ...
poincareSection = struct('endPosition',{},'numPoints',{},'orbitMaxLength',{});
...
poincareSection(1).endPosition = [3.3,-32.1;3.7,-31.6];
poincareSection(2).endPosition = [1.3,-30.9;1.9,-31.1];

% Number of orbit seed points along each Poincare section
[poincareSection.numPoints] = deal(100);

% Set maximum orbit length to twice the expected circumference
nPoincareSection = numel(poincareSection);
for i = 1:nPoincareSection
    rOrbit = hypot(diff(poincareSection(i).endPosition(:,1)),diff(poincareSection(i).endPosition(:,2)));
    poincareSection(i).orbitMaxLength = 2*(2*pi*rOrbit);
end

for lambda = lambdaRange
...
    [shearline.etaPos,shearline.etaNeg] = lambda_line(cgEigenvector,cgEigenvalue,lambda);
    closedLambdaLineCandidate = poincare_closed_orbit_multi(domain,resolution,shearline,poincareSection,'odeSolverOptions',lambdaLineOdeSolverOptions,'showGraph',showGraph);
...
end
...
% Hyperbolic shrinkline LCSs
shrinklineLcs = seed_curves_from_lambda_max(shrinklineLocalMaxDistance,shrinklineMaxLength,cgEigenvalue(:,2),cgEigenvector(:,1:2),domain,resolution,'odeSolverOptions',shrinklineOdeSolverOptions);
...
% Hyperbolic stretchline LCSs
stretchlineLcs = seed_curves_from_lambda_max(stretchlineLocalMaxDistance,stretchlineMaxLength,-cgEigenvalue(:,1),cgEigenvector(:,3:4),domain,resolution,'odeSolverOptions',stretchlineOdeSolverOptions);
\end{lstlisting}

We choose the integration time as $T=30$ days (Listing \ref{l:griddedInterp},
l.\,4)\textbf{,} which is larger than the eddy turnover time in this
region. The resolution of the main computational grid for initial
conditions is set to $400\times400$ (l.\,3). This corresponds to
a resolution of roughly 0.015\degree\, and gives good results. With
this choice, the resolution of the tracer grid is 15 times higher
than the resolution of the velocity field. The flow is integrated
and the Cauchy-Green strain tensor is computed by the function \inputencoding{latin9}\lstinline!eig_cgStrain!\inputencoding{utf8}
(l.\,35). Incompressibility of the flow is enforced (l.\,12), the
auxiliary grid distance is set to 1\% of the main grid distance (l.\,17),
and eigenvalues are computed from the auxiliary grid (l.\,16). Elliptic
LCSs are computed in line 54, after the Poincare sections have been
set (ll.\,41-42) and the $\eta_{\pm}^{\lambda}$ fields have been
defined (l.\,53). $\lambda$-values are varied over a range of $[0.90,1.10]$
with a step of $0.02$ (l.\,20). Hyperbolic LCSs are computed in
lines 57 and 59.

\begin{figure}
\centering
\includegraphics[width=0.9\textwidth]{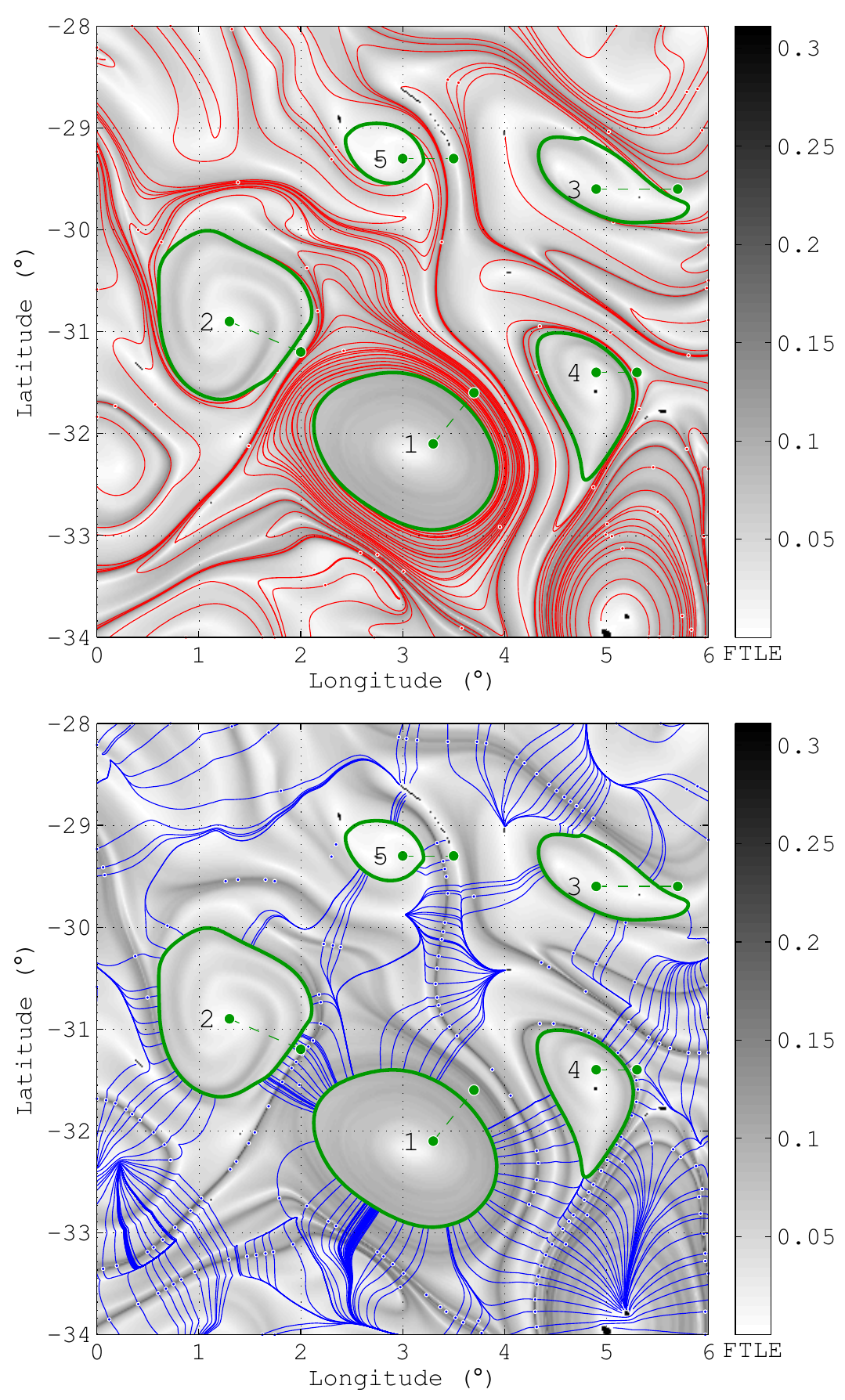}
\protect\caption{(Top) Elliptic LCSs (green, $\lambda=$ 1.00, 1.08, 0.94, 0.90, 1.06)
and repelling LCSs (red) in the ocean data set. Background gray-scale
is the finite time Lyapunov exponent. Dashed green lines are Poincare
sections. White dots indicate $\lambda_{2}$ local maxima used as
initial positions for shrinkline integration. (Bottom) Elliptic LCSs
(green) and attracting LCSs (blue).}

\label{f:ocean dataset hyperbolic shear lcs details strainline} 
\end{figure}

Figure~\ref{f:ocean dataset hyperbolic shear lcs details strainline}
shows the positions of elliptic and hyperbolic LCSs on 22 November
2006, which is the same time instant analyzed in~\citep{haller13:_coher_lagran,beron-vera13:_objec_agulh}.
Here the integration time is chosen as $T=30$ days, as opposed to
90 days in the cited references, to avoid the length tangling of hyperbolic
LCSs. Our analysis via LCS Tool reveals five coherent eddy boundaries.
The large coherent eddy at $(3,-32)$ has a non-stretching boundary,
i.e., $\lambda=1.00$. It corresponds to eddy \#2 in Figure~3 of
\citep{beron-vera13:_objec_agulh}. Additionally, four further smaller
coherent eddy cores are found. They do not stay coherent over a time
of 90 days. All closed orbits are stable under increased spatial resolution
for the Cauchy-Green strain tensor field. Repelling LCSs (red) and
attracting LCSs (blue) are superimposed over the plot of FTLE values.
The hyperbolic LCSs determine the deformation of the fluid in-between
coherent eddy cores.

\clearpage{}

\section{Conclusions}

We have described a computational toolbox, LCS Tool, that implements
recent variational results for Lagrangian Coherent Structures (LCSs)
in two-dimensional unsteady flows. We have also demonstrated the performance
of LCS Tool on two analytic flow models and a geophysical velocity dataset. The publicly available software library producing these
results enables the exploration of variational LCS methods without
assuming a detailed knowledge of geodesic LCS theory. 

LCS Tool leverages the capabilities of MATLAB extensively. For FTLE-based
extraction of LCSs, computational performance has received considerable
attention \citep{conti12:_gpu_apu_finit_time_lyapun_expon,miron12:_anisot_lagran_coher_struc}.
We hope to see similar computational advances for the publicly available
LCS Tool platform. Optimizing computational performance will likely
aid the application of variational LCS methods to large-scale real-time
forecasting applications, such as the tracking of environmental contaminants
\citep{Olascoaga2012}.

We also hope that LCS Tool will serve as a foundation for the numerical
implementation of recent theoretical advances. These include the geodesic
theory of parabolic LCSs (jet cores) \citep{Farazmand2014}
and the variational theory of hyperbolic and elliptic LCSs for three-dimensional
flows \citep{Blazevski2014}.

\subsection*{Acknowledgments}

The altimeter products used in this work are produced by SSALTO/DUACS
and distributed by AVISO, with support from CNES \url{http://www.aviso.oceanobs.com/duacs}

\bibliographystyle{plainnat}
\bibliography{main}

\end{document}